\documentclass[11pt,a4paper]{article}

\usepackage{amsmath}
\usepackage{mathtools}
\usepackage{nccmath}
\usepackage{amsthm}
\usepackage{amssymb}
\usepackage{amsfonts}
\usepackage{anyfontsize}
\usepackage{mathrsfs}
\usepackage{bbm}
\usepackage{graphicx}
\usepackage{caption}
\usepackage{cite}
\usepackage{bookmark,hyperref}
\usepackage{12many}
\usepackage{parskip}
\usepackage{setspace}
\usepackage{multirow}
\usepackage{arydshln}
\usepackage{anyfontsize}
\usepackage{changepage}

\newcommand\blfootnote[1]{%
  \begingroup
  \renewcommand\thefootnote{}\footnote{#1}%
  \addtocounter{footnote}{-1}%
  \endgroup}

\newcommand{\SLtwoZ}{\mathrm{SL}(2,\mathbb{Z})}
\newcommand{\SLtwoR}{\mathrm{SL}(2,\mathbb{R})}

\DeclareMathOperator{\diag}{diag}
\numberwithin{equation}{section}
\usepackage{fullpage}

\begin{document}
\begin{titlepage}
 \thispagestyle{empty}
 \begin{flushright}

 \end{flushright}

 \vspace{30pt}

 \begin{center}
     
  {\fontsize{18}{22} \bf {Towards a String Realization of the Dark Dimension via T-folds}}

     \vspace{30pt}
{\fontsize{13}{16}\selectfont {Guo-En Nian and Stefan Vandoren}} \\[10mm]

{\small\it
Institute for Theoretical Physics  \\
Utrecht University, 3508 TD Utrecht, The Netherlands \\[3mm]}

\vspace{1.5cm}

{\bf Abstract}

\vspace{0.3cm}
   
\begin{adjustwidth}{12pt}{12pt}
In this work, we explore the feasibility of realizing the Dark Dimension Scenario through T-fold compactifications on $T^5 \times S^1$, where the base of the internal space $S^1$ naturally acts as the requisite mesoscopic extra dimension. Utilizing Scherk--Schwarz reduction from 5D to 4D, and applying duality twists from both elliptic and parabolic conjugacy classes of subgroups of the T-duality group associated with $T^5$, we stabilize the volume of $T^5$ and several other moduli. This approach generates a potential characterized by two runaway directions, one aligned with the Scherk--Schwarz radion. Further stabilization efforts yield a decreasing potential trajectory demanded by the Dark Dimension Scenario.

\end{adjustwidth}

\end{center}

\vspace{20pt}

\blfootnote{g.nian@uu.nl \quad s.j.g.vandoren@uu.nl}

\noindent

\end{titlepage}

\begin{spacing}{1.15}
\tableofcontents
\end{spacing}

\section{Introduction}
There is a longstanding history of connecting string theory with cosmology, particularly in constructing an effective theory that accounts for the small and positive dark energy observed in our Universe. Two main approaches are considered, both of which are challenging: one involves realizing a metastable de Sitter vacuum, and the other involves constructing a cosmological model with dynamical dark energy. In the latter case, the dark energy is characterized by a decreasing scalar potential. Recent observations by DESI \cite{DESI:2024mwx,DESI:2024hhd} suggest that the dark energy in our Universe might be time-dependent, lending support to the second approach, though previous astronomical observations prefer a cosmological constant. 

Meanwhile, the effective theories consistent with quantum gravity satisfy some universal criteria. These criteria are called Swampland conjectures, and distinguish the effective cosmological models and particle physics in the Landscape or the Swampland. By relating Swampland conjectures to certain observational data of our Universe, a particular corner of the string landscape, known as the Dark Dimension Scenario \cite{Montero:2022prj}, emerges. This scenario connects the neutrino mass scale and the dark energy scale, with the smallness of both corresponding to an asymptotic region of the field space. The Dark Dimension Scenario requires the existence of exactly one mesoscopic extra dimension, referred to as the dark dimension, and the scalar potential behaves as $V\propto m_{\text{KK}}^4$, where $m_{\text{KK}}$ is the Kaluza--Klein mass scale of this dark dimension. We will provide a concise introduction to the Dark Dimension in Sec.~\ref{sec2}. The scenario also predicts the sterile neutrino scale and the strong gravity scale, which are beyond the scope of this paper and will not be discussed. There have been attempts to find specific models satisfying the Dark Dimension Scenario, such as in \cite{Blumenhagen:2022zzw,Danielsson:2022lsl,Cui:2023wzo,Anchordoqui:2023oqm,Heckman:2024trz,Basile:2024lcz}. For instance, \cite{Blumenhagen:2022zzw} tries to realize this scenario near a warped throat, but it is challenging to obtain the mass of the Kaluza--Klein tower along the dark dimension. In \cite{Anchordoqui:2023oqm}, the dark dimension potential behavior is realized using an $STU$-model under Scherk--Schwarz reduction from $\mathcal{N}=1$ supergravity. And \cite{Basile:2024lcz} determines the Dark Dimension potential from the Casimir energy of the worldsheet torus partition function in infinite-distance limits.

In addition, non-geometric compactifications, which form a significant subset of string compactifications, have been studied for over two decades. In non-geometric compactifications, the internal space is a fiber bundle whose transition functions combine diffeomorphisms with duality transformations. A key example is the T-fold, which we will discuss in this work, where the duality involved is T-duality. The effective field theory of these compactifications is obtained via the Scherk--Schwarz mechanism, and T-duality introduces twists in the periodicity conditions of fields. Depending on the twist from the duality group, the internal space is altered, corresponding to the generation of fluxes over the non-geometric space. These fluxes can be understood as arising from T-duality transformations of the $H$-fluxes.

Previous investigations into non-geometric string realizations of de Sitter vacua have shown that such string theories are inconsistent with several no-go theorems established \cite{Plauschinn:2020ram,Prieto:2024shz}. In this work, we focus instead on non-geometric string compactifications with dynamical dark energy and attempt to realize the Dark Dimension scenario.

The paper is organized as follows. In Sec.~\ref{sec2}, we provide an introduction to the Dark Dimension scenario, focusing particularly on the behavior of dynamical dark energy. In Sec.~\ref{sec3}, we introduce the Scherk--Schwarz reduction and derive its potential. We consider nine-dimensional type IIB supergravity as an example of Scherk--Schwarz stringy uplift, which has the $\SLtwoZ$ S-duality group as the twist group. In Sec.~\ref{sec4}, we present a toy model as a precursor to the string realization of the Dark Dimension. This model involves the pure gravity theory of a type II T-fold string compactification from 10D to 7D, with $T^2$ T-duality $\mathrm{O}(2,2;\mathbb{Z})$ as the twist group. In Sec.~\ref{sec5}, we realize the scalar potential behavior required by the Dark Dimension scenario in four dimensions, via $T^5\times S^1$ T-fold compactification with $T^5$ T-duality as the twist group.

Throughout this paper, we set the four-dimensional Planck mass $M_{\text{Pl}}=1$.

\section{Dark Dimension Scenario}\label{sec2}
In this section, we provide a brief overview of the Dark Dimension Scenario proposed in \cite{Montero:2022prj}. This scenario offers a compelling framework that connects string phenomenology with observable cosmological parameters.

The theoretical foundation of the scenario is built upon two key Swampland criteria. The first criterion is the (A)dS Distance Conjecture \cite{L_st_2019}. This conjecture posits that, when approaching the infinity of the field space, the distance over the field space scales proportionally to $\log\left(1/|\Lambda|\right)$, where $\Lambda$ is the cosmological ``constant". Meanwhile, in the context of string compactifications, there exists a tower of states with the smallest mass scale near the infinity of the field space. The mass scale of the tower decreases exponentially with the geodesic distance. This is proposed as the Swampland Distance Conjecture (SDC) \cite{Ooguri:2006in}. Consequently, when $\Lambda\to 0$, the mass scale $m$ of the tower of states adheres to the following relation
\begin{equation}\label{dSDC}
m = \lambda^{-1} |\Lambda|^\alpha, \quad \alpha\sim\mathcal{O}(1)\,.
\end{equation}
For $\Lambda >0$, the Higuchi bound \cite{Higuchi:1986py} imposes an upper limit on the exponent $\alpha$, specifically $\alpha\leq\frac{1}{2}$. This bound ensures that the classical contribution to the potential does not exceed $V\sim m^2$. Additionally, there exists a lower bound on $\alpha$, given by $\alpha \geq \frac{1}{d}$, where $d$ denotes the number of external dimensions. This lower bound arises from the 1-loop Casimir energy contribution to the potential, $V_{\text{Casimir}} \sim m^d$, in scenarios where there are more massless fermions than massless bosons. When $\alpha$ attains its lower bound, the parameter $\lambda$ in equation \eqref{dSDC} is expected to satisfy the relation $m^{1/2} \lesssim \lambda^4 \lesssim 1$. This implies that the cosmological constant $\Lambda$ must lie within the range $m^{9/2} \lesssim \Lambda \lesssim m^4$ in four dimensions. Furthermore, if the potential is predominantly influenced by the Casimir corrections asymptotically, the parameter $\lambda$ should approach its upper bound. The values of the parameters $\alpha$ and $\lambda$ are crucial for aligning the theoretical predictions with realistic observational data.

The second essential Swampland conjecture is the Emergent String Conjecture (ESC) \cite{Lee:2019wij}. Based on the SDC, the ESC further states that, approaching the asymptotic limits of the field space, there are only two possible types of towers of states that can emerge. These towers must consist of either:
\begin{itemize}
    \item A tower of string excited states, or
    \item a Kaluza--Klein tower associated with decompactification.
\end{itemize}
To determine which of these cases corresponds to the mass scale described in equation \eqref{dSDC}, it is necessary to consider observational and experimental data from our universe. In the first case, the mass scale corresponds to the string scale, at which the local EFT ceases to be valid; in the second case, the presence of decompactifying extra dimensions influences gravity propagation. Experimental constraints from torsion balance experiments, which measure deviations in Newton's gravitational force, are particularly relevant for detecting the smallest possible mass scales. The most recent findings \cite{Lee:2020zjt} impose that
\begin{equation}\label{m}
m^{-1}\lesssim 30\ \mu\text{m},\quad \text{i.e.} \quad m \gtrsim 6.6\ \mathrm{meV}\,.
\end{equation}
Interestingly, this mass scale coincides with the neutrino scale. In addition, the observed value of the cosmological constant is $\Lambda\sim 10^{-122}$, which is too small and implies $\Lambda^{1/4}=2.31\ \mathrm{meV} \lesssim m$. Therefore, within the framework of \eqref{dSDC}, we anticipate the parameters to satisfy:
\begin{equation}
\alpha = \frac{1}{4},\quad\text{and}\quad \lambda\sim 10^{-1}\,.
\end{equation} 
If this minimal tower with mass scale \eqref{m} corresponds to the first case of the ESC, there would be no valid EFT beyond $m$, contradicting our current understanding of the universe. Hence, it is more plausible that the mass scale $m$ in \eqref{dSDC} represents a Kaluza--Klein decompactification scale, with neutrinos acting as Kaluza--Klein spinors.

Further constraints on the Kaluza--Klein scale arise from observations of KK gravitons in the neutron star surrounding cloud \cite{10.1093/ptep/ptaa104,PhysRevD.67.125008}. Specifically, if there is only one relatively large extra dimension and the other extra dimensions are negligible, the upper bound on the KK scale is $m_{\text{KK}}^{-1}<44\ \mu\text{m}$; if there are two relatively large extra dimensions, the upper bound becomes $m_{\text{KK}}^{-1}< 1.6\times 10^{-4}\ \mu\text{m}$; the bound becomes even smaller for more large extra dimensions. Only the first scenario, involving a single large extra dimension, is consistent with the constraint in equation \eqref{m}. This single large extra dimension is referred to as the Dark Dimension in \cite{Montero:2022prj}. In this context, the scalar potential is given by:
\begin{equation}
V = \Lambda \propto m_{\text{KK}}^4\,,
\end{equation}
where $m_{\text{KK}}$ denotes the Kaluza--Klein scale associated with the Dark Dimension. The seminal paper \cite{Montero:2022prj} emphasizes the potential with $V\sim m^4$ interpreted as a 1-loop Casimir correction. The computation of the Casimir correction requires a worldsheet torus partition function, which is only known for some familiar models. In our work, we explore a model in which the Scherk--Schwarz potential exhibits the necessary behavior to align with the Dark Dimension Scenario.

\section{Scherk--Schwarz reduction}\label{sec3}
We begin by considering a Scherk--Schwarz reduction \cite{Scherk:1978ta,Scherk:1979zr} from $d+1$ to $d$ dimensions, with coordinates $x^{\hat{\mu}}=\left\{x^\mu,y\right\},\,\mu=0,1,\cdots,d-1$. The $d$-th dimension compactifies to a circle, with periodicity condition $y \sim y +2\pi R$. The radius of the circle $\varrho$ is modulated by the radion field $\phi$, such that
\begin{equation}
\varrho = R e^{\beta\phi}\,,
\end{equation}
where $\beta$ is a real constant to be determined later. 

Assuming the theory inherits a global symmetry $G$ from higher dimensions, a field $\hat{\psi}$ in the fundamental representation transforms under the Scherk--Schwarz reduction can be ansated as
\begin{equation}\label{hatpsi}
\hat{\psi}\left(x^{\mu},\,y\right) = \exp\left(\frac{My}{2\pi R}\right) \psi\left(x^\mu\right), \quad M\in \mathfrak{g}\,.
\end{equation}
In many cases, the scalar fields parametrize a coset space $G/H$ with $H$ the maximal compact subgroup of $G$, and can be arranged into a matrix field $\hat{\mathcal{H}}$, which transforms as the adjoint representation of $G$ and expands as
\begin{equation}\label{hatH}
\hat{\mathcal{H}}\left(x^{\mu},\,y\right) = \exp\left(\frac{My}{2\pi R}\right) \mathcal{H}\left(x^\mu\right)\exp\left(\frac{M^T y}{2\pi R}\right)\,.
\end{equation}
The $d+1$-dimensional bosonic action involving a $G$-matrix scalar field in the Einstein frame is
\begin{equation}
S = \int\mathrm{d}^{d+1}x \sqrt{-\hat{g}_{(d+1)}} \frac{1}{2\kappa_{(d+1)}^2} \left[R_{(d+1)} + \operatorname{Tr}\left(\hat{g}^{\hat{\mu}\hat{\nu}}\partial_{\hat{\mu}}\hat{\mathcal{H}}^{-1} \partial_{\hat{\nu}}\hat{\mathcal{H}}\right) \right]\,.
\end{equation}
Upon reduction to $d$ dimensions, the metric $\hat{g}_{\hat{\mu} \hat{\nu}}$ in $d+1$ dimensions transforms as:
\begin{equation}
\hat{g}_{\hat{\mu} \hat{\nu}}=\left(\begin{array}{cc}
e^{2 \alpha \phi} g_{\mu \nu} & \\
 & e^{2 \beta \phi}
\end{array}\right)\,,
\end{equation}
where $g_{\mu\nu}$ is the $d$-dimensional metric, and $\phi(x^\mu)$ denotes the radion field. We only focus on the scalar potential and ignore graviphoton contributions for this analysis. To decouple the radion and the curvature, and normalize the kinetic term of the radion, it is derived that
\begin{equation}
\alpha=-\sqrt{\frac{1}{2(d-1)(d-2)}}\,, \quad \beta=-(d-2) \alpha\,.
\end{equation}
Especially for $d=4$, the coefficients simplify to
\begin{equation}
\alpha = -\frac{1}{2\sqrt{3}}\,,\quad \beta = \frac{1}{\sqrt{3}}\,.
\end{equation}
Following the reduction to $d$ dimensions, the relevant dimensional equations and transformations are
\begin{eqnarray}
\int\mathrm{d}^{d+1}x &=&  2\pi R\int\mathrm{d}^{d}x\,,\\
\sqrt{-g_{(d+1)}} &=& \sqrt{-g_{(d)}} e^{(d\alpha+\beta)\phi} = \sqrt{-g_{(d)}} e^{2\alpha\phi}\,,\\
\frac{1}{2\kappa_{(d+1)}^2} &=& \frac{1}{2\kappa_{(d)}^2 2\pi R}\,,\\
R_{(d+1)} &=& e^{-2\alpha\phi} R_{(d)} - \frac{1}{2}e^{-2\alpha\phi}g^{\mu\nu}\partial_{\mu}\phi \partial_{\nu}\phi + \text{bdr.}\,,\\
\operatorname{Tr}\left(\hat{g}^{\hat{\mu}\hat{\nu}}\partial_{\hat{\mu}}\hat{\mathcal{H}}^{-1} \partial_{\hat{\nu}}\hat{\mathcal{H}}\right) &=& e^{-2\alpha\phi}\operatorname{Tr}\left(g^{\mu\nu}\partial_{\mu}\mathcal{H}^{-1} \partial_{\nu}\mathcal{H}\right) - 2e^{-2\beta\phi}\operatorname{Tr}\left(\frac{M^2}{(2\pi R)^2}+ \frac{M^T}{2\pi R}\mathcal{H}^{-1}\frac{M}{2\pi R}\mathcal{H}\right).\quad\quad\quad
\end{eqnarray}
Thus, the $d$-dimensional action within the Einstein frame is formulated as 
\begin{equation}
S = \int\mathrm{d}^{d}x \sqrt{-g_{(d)}} \frac{1}{2\kappa_{(d)}^2} \left[ R_{(d)} - \frac{1}{2}g^{\mu\nu}\partial_{\mu}\phi \partial_{\nu}\phi + \operatorname{Tr}\left(g^{\mu\nu}\partial_{\mu}\mathcal{H}^{-1} \partial_{\nu}\mathcal{H}\right) - V \right],
\end{equation}
where the potential $V$ is
\begin{equation}\label{V}
V=2e^{2(\alpha-\beta)\phi}\operatorname{Tr}\left(\frac{M^2}{(2\pi R)^2}+ \frac{M^T}{2\pi R}\mathcal{H}^{-1}\frac{M}{2\pi R}\mathcal{H}\right).
\end{equation}

Note that $\mathcal{H}\in G/H$ and for $M$ a real matrix, this potential is non-negative definite. The potential becomes zero specifically when $M$ conjugates to a rotation generator, i.e. there is a constant matrix $S\in G$, such that $\widetilde{M}=SMS^{-1}$ and $\widetilde{M}= -\widetilde{M}^T$. In this case, there can be a Minkowski vacuum. For example, the elliptic conjugacy classes of $G=\SLtwoR$ give a classical Minkowski vacuum at $\tau = i$ (see e.g. \cite{Dabholkar_2003}), which will be further discussed in the next subsection.

In string theory, duality groups provide a natural candidate for the non-Abelian global symmetries utilized in Scherk--Schwarz reductions. When the global symmetry group $G$ is a duality group, the corresponding group element $\mathcal{M}=e^M\in G$ is known as a duality twist, and fields like those in \eqref{hatpsi} and \eqref{hatH} acquire twisted boundary conditions characterized by a monodromy matrix $M\in \mathfrak{g}$. Such a twist effectively deforms the geometry of the internal space, an effect that is equivalent to introducing background fluxes. In Section~\ref{sec4}, we demonstrate this principle by deducing the NS-NS fluxes from T-duality twists and subsequently parametrizing the potential \eqref{V} in terms of the resulting flux numbers.

We aim to align this potential with the required potential proportional to the Kaluza--Klein scale in the Dark Dimension Scenario. The Kaluza--Klein scale in string frame is known as
\begin{equation}
m_{\text{KK}}^{(\text{S})} = \frac{1}{\varrho} =\frac{1}{R}e^{-\beta\phi}\,.
\end{equation}
To convert to the Einstein frame, the square of the mass scale should be multiplied by the metric factor $e^{2\alpha\phi}$, such that
\begin{equation}
m_{\text{KK}}^2 = \frac{1}{R^2}e^{2(\alpha-\beta)\phi}\,.
\end{equation}
Notably, $m_{\text{KK}}^2$ is proportional to the part outside the trace parentheses of \eqref{V}. However, as highlighted previously, the Dark Dimension Scenario necessitates that $V \propto m_{\text{KK}}^4$. This suggests that the contributions inside the trace parentheses to the potential should influence the radion's exponent at least equally as the outside. In the next sections, we will demonstrate how it is naturally realized in our model, via a specific parabolic duality twist of T-fold constructions. Without loss of generality, from now on we take $2\pi R =1$ for convenience. 

\subsection{Potential with $\SLtwoR$ and $\SLtwoZ$ monodromy}
The stringy uplift of the Scherk--Schwarz mechanism typically incorporates a duality twist within a duality group $G(\mathbb{Z})$. A straightforward example of this is provided by compactifying type IIB supergravity on a circle to yield a nine-dimensional theory, with S-duality group $\mathrm{SL}(2,\mathbb{Z})$ acting as the twist group. This group operates on the axio-dilaton field $\tau=\tau_1+i\tau_2$, which in the adjoint representation is expressed as
\begin{equation}\label{Htau}
\mathcal{H}(\tau) = \frac{1}{\tau_2}\begin{pmatrix}
|\tau|^2 & \tau_1 \\
\tau_1 & 1
\end{pmatrix}.
\end{equation}
The Scherk--Schwarz potential \eqref{V} is intrinsically characterized by the conjugacy classes of the twist group. To start with, we first examine the conjugacy classes of the classical S-duality group $G=\mathrm{SL}(2,\mathbb{R})$ which corresponds to the truncated theory with only massless states. This group comprises three kinds of conjugacy classes, which are called: parabolic, if $|\operatorname{Tr}\mathcal{M}|=2$; elliptic, if $|\operatorname{Tr}\mathcal{M}|<2$; and hyperbolic, if $|\operatorname{Tr}\mathcal{M}|>2$.
Corresponding to these classes, the monodromy matrices can be represented as \cite{Hull:1998vy,Dabholkar_2003}
\begin{equation}
\mathcal{M}_p=\left(\begin{array}{cc}
1 & m \\
0 & 1
\end{array}\right), \quad
\mathcal{M}_e=\left(\begin{array}{cc}
\cos m & \sin m \\
-\sin m & \cos m
\end{array}\right), \quad \mathcal{M}_h=\left(\begin{array}{cc}
e^m & 0 \\
0 & e^{-m}
\end{array}\right) ,
\end{equation}
where $m$ is called the mass parameter. These matrices correspond to shift, rotation, and boost transformations for the parabolic, elliptic, and hyperbolic conjugacy classes, respectively. The generators $M$ associated with these transformations are given by:
\begin{equation}\label{Mpeh}
M_p=\left(\begin{array}{cc}
0 & m \\
0 & 0
\end{array}\right), \quad
M_e=\left(\begin{array}{cc}
0 & m \\
-m & 0
\end{array}\right), \quad M_h=\left(\begin{array}{cc}
m & 0 \\
0 & -m
\end{array}\right) .
\end{equation}

Utilizing equation \eqref{Htau} and the generator matrices in \eqref{Mpeh}, we can compute the Scherk--Schwarz potential \eqref{V} corresponding to monodromies in different conjugacy classes as follows:
\begin{itemize}
\item Parabolic monodromy: The potential becomes
\begin{equation}\label{Vp}
V_p = 2m^2 e^{-a\phi} \frac{1}{\tau_2^2}\,, \quad a\equiv 2(\beta-\alpha)\,.
\end{equation}
This form of the potential indicates that $\tau_1$ remains a flat direction, while both $\phi$ and $\tau_2$ tend to run away towards $+\infty$. 
\item Elliptic monodromy: For the elliptic monodromy $\mathcal{M}_e$ of $\SLtwoR$, the potential is given by
\begin{equation}\label{Ve}
V_e = 2m^2 e^{-a\phi} \left( \frac{|\tau|^4 + 2\tau_1^2+1}{\tau_2^2} -2\right).
\end{equation}
This potential is non-negative and possesses a minimum at $\tau = i$, corresponding to a Minkowski vacuum.
\item Hyperbolic monodromy: The potential is
\begin{equation}
V_h = 8m^2 e^{-a\phi}\left( \frac{\tau_1^2}{\tau_2^2} +1 \right).
\end{equation}
In this case, $\tau_1$ is stabilized at $\tau_1=0$ and $\tau_2$ is a flat direction, such that at $\tau_1=0$, the potential simplifies to
\begin{equation}\label{Vhs}
V_{h,\text{ stab.}}= 8 m^2 e^{-a\phi}\,.
\end{equation}
\end{itemize}
The rationale for introducing potentials with monodromies corresponding to the conjugacy classes of $\SLtwoR$ lies in the fact that $\SLtwoZ$ monodromies are conjugate to these classes. When incorporating massive states into the theory, the monodromy matrices $\mathcal{M}$ must belong to $\SLtwoZ$. In this context, the parabolic conjugacy classes retain the same form as in $\SLtwoR$, with the mass parameter $m\in \mathbb{Z}$; the elliptic and hyperbolic classes correspond to integer conjugations of $\mathcal{M}_e$ and $\mathcal{M}_h$ respectively. Notably, there exist four rotational $\SLtwoZ$ conjugacy classes with representatives \cite{DeWolfe:1998pr,Dabholkar_2003}
\begin{equation}
\mathcal{M}_2=\left(\begin{array}{cc}
-1 & 0 \\
0 & -1
\end{array}\right), \quad \mathcal{M}_3=\left(\begin{array}{cc}
0 & 1 \\
-1 & -1
\end{array}\right), \quad \mathcal{M}_4=\left(\begin{array}{cc}
0 & 1 \\
-1 & 0
\end{array}\right), \quad \mathcal{M}_6=\left(\begin{array}{cc}
1 & 1 \\
-1 & 0
\end{array}\right),
\end{equation}
with corresponding generators
\begin{equation}\label{M2346}
M_2=\pi\left(\begin{array}{cc}
0 & 1 \\
-1 & 0
\end{array}\right), \, M_3=\frac{2\pi}{3\sqrt{3}}\left(\begin{array}{cc}
1 & 2 \\
-2 & 1
\end{array}\right), \, M_4=\frac{\pi}{2}\left(\begin{array}{cc}
0 & 1 \\
-1 & 0
\end{array}\right), \, M_6=\frac{\pi}{3\sqrt{3}}\left(\begin{array}{cc}
1 & 2 \\
-2 & 1
\end{array}\right).
\end{equation}
These matrices correspond to the $\mathbb{Z}_2,\,\mathbb{Z}_3,\,\mathbb{Z}_4,\,\mathbb{Z}_6$ subgroups of $\SLtwoZ$ respectively. $\mathcal{M}_3,\,\mathcal{M}_4,\,\mathcal{M}_6$ are elliptic, while $\mathcal{M}_2$ represents the $m\to\pi$ limit of $\mathcal{M}_e$ in $\SLtwoR$. Furthermore, $\mathcal{M}_3$ and $\mathcal{M}_6$ are $\SLtwoR$-conjugate to $\mathbb{Z}_3$ and $\mathbb{Z}_6$ elements of $\mathcal{M}_e$. For example,
    \begin{equation}
        \mathcal{M}_6 = \left(\begin{array}{cc}
\sqrt{\frac{2}{\sqrt{3}}} & 0 \\
-\sqrt{\frac{1}{2\sqrt{3}}} & \sqrt{\frac{\sqrt{3}}{2}}
\end{array}\right) \mathcal{M}_e\left(\frac{\pi}{3}\right) \left(\begin{array}{cc}
\sqrt{\frac{2}{\sqrt{3}}} & 0 \\
-\sqrt{\frac{1}{2\sqrt{3}}} & \sqrt{\frac{\sqrt{3}}{2}}
\end{array}\right)^{-1}.
\end{equation}
Incorporating these conjugacy class representatives into the potential expression \eqref{V} allows us to derive the corresponding potentials. It is straightforward to verify that, for $\mathcal{M}_2,\,\mathcal{M}_4$ the potential exhibits the same minimum as in equation \eqref{Ve}. For $\mathcal{M}_3$ and $\mathcal{M}_6$, the potentials are equivalent to \eqref{Ve} up to conjugations, with shifted locations for the minima. It can be calculated that the Minkowski minimum is at $\tau = \exp{\frac{2\pi i}{3}}$.

The hyperbolic conjugacy classes of $\SLtwoZ$ are a bit complicated. In the following section, we will explore hyperbolic conjugacy classes in greater detail in terms of T-fold fluxes. Nevertheless, as indicated in equation \eqref{Vhs}, hyperbolic conjugacy classes contribute trivially to the potential, thereby precluding the derivation of the Dark Dimension ratio from these classes.

\section{$T^2\times S^1$ T-fold as a toy model}\label{sec4}
This section discusses the feasibility of realizing the Dark Dimension Scenario through potentials derived from T-fold compactifications, where the twist group is the T-duality group. To facilitate comprehension of our construction, we begin by presenting a toy model. We consider the compactification of 10-dimensional type II superstring theory down to 8 dimensions via a $T^2$ reduction, focusing exclusively on the pure gravity spectrum. Subsequently, we perform a Scherk--Schwarz reduction over an additional $S^1$, utilizing a twist matrix $\mathcal{M}=e^M$ that belongs to the T-duality group associated with $T^2$. 

Denote the coordinates of $T^2$ as $\{z_1,\,z_2\}$, and the $S^1$-coordinate as $y$. The compactification over $T^2$ introduces one complexified K\"ahler modulus $\rho$ and one complex structure modulus $\tau$, defined as
\begin{equation} \label{taurho}
\tau = \tau_1+i\tau_2= \frac{g_{12}}{g_{22}}+i\frac{\sqrt{g}}{g_{22}}\,,\quad \rho = \rho_1+i\rho_2 = \frac{1}{\alpha'}\left(b_{12} + i\sqrt{g}\right),
\end{equation}
which parameterize the moduli space
\begin{equation}
\mathrm{O}(2,2;\mathbb{Z})\backslash \mathrm{O}(2,2;\mathbb{R})/(U(1)\times U(1))\,.
\end{equation}
The corresponding T-duality group is given by
\begin{equation}\label{SO22}
G(\mathbb{Z})=\mathrm{O}(2,2;\mathbb{Z}) \cong \SLtwoZ\times\SLtwoZ\times \mathbb{Z}_2\ltimes \mathbb{Z}_2\,.
\end{equation}
As discussed in the previous subsection, we considered $\SLtwoR$ and certain $\SLtwoZ$ monodromies. From \eqref{SO22} it is evident that each $\SLtwoZ$ corresponds to one of the two moduli, representing half of the T-duality group. Analogous to \eqref{Htau}, these two moduli can be expressed in the representations of $\SLtwoR$ as
\begin{equation}\label{Htaurho}
\mathcal{H}_\tau = \frac{1}{\tau_2}\left(\begin{array}{cc}
|\tau|^2 & \tau_1 \\
\tau_1 & 1
\end{array}\right),\quad \mathcal{H}_\rho = \frac{1}{\rho_2}\left(\begin{array}{cc}
|\rho|^2 & \rho_1 \\
\rho_1 & 1
\end{array}\right).
\end{equation}

Besides these geometric moduli, the $T^2$ compactification yields two Ramond-Ramond real scalars, which transform as spinors under T-duality \cite{Bergshoeff:1995as,Hassan_2000}, and one lower-dimensional dilaton, which is invariant under T-duality. In our analysis, we concentrate on T-fold constructions, which cannot generate potentials involving these additional fields. Hence we truncate these fields to ensure that the remaining scalar fields parameterize the quotient group $G/H$. The truncated fields, together with the Kähler modulus, transform under the adjoint representation of $\mathrm{SL}(3,\mathbb{R})$, corresponding to a part of the U-duality group. A detailed discussion of such a U-fold compactification is provided in Appendix~\ref{AppA}.

\subsection{Monodromy classification in terms of fluxes}\label{hq}
In this subsection, we examine the classification of monodromies in terms of fluxes within the framework of T-fold compactifications. Specifically, we explore how fluxes arise from duality twists in the T-duality group and their implications for the Dark Dimension Scenario.

Generically, the NS-NS moduli of a torus $T^n$ excluding the dilaton can be represented as the generalized metric
\begin{equation}\label{H}
\mathcal{H}=\left(\begin{array}{cc}
\frac{1}{\alpha^{\prime}}\left(g-b g^{-1} b\right) & b g^{-1} \\
-g^{-1} b & \alpha^{\prime} g^{-1}
\end{array}\right) \in \mathrm{O}(n,n)\,,
\end{equation}
where $g$ is the metric and $b$ is the Kalb--Ramond two-form field. For a two-dimensional torus $T^2$, comparing with \eqref{taurho}, the moduli matrix under this representation becomes
\begin{equation}\label{H2}
\mathcal{H}=\frac{1}{\rho_2 \tau_2}\left(\begin{array}{cccc}
|\rho|^2|\tau|^2 & |\rho|^2 \tau_1 & -\rho_1 \tau_1 & \rho_1|\tau|^2 \\
|\rho|^2 \tau_1 & |\rho|^2 & -\rho_1 & \rho_1 \tau_1 \\
-\rho_1 \tau_1 & -\rho_1 & 1 & -\tau_1 \\
\rho_1|\tau|^2 & \rho_1 \tau_1 & -\tau_1 & |\tau|^2
\end{array}\right).
\end{equation}
The element $\mathcal{M}$ of the T-duality group $\mathrm{O}(n,n;\mathbb{Z})$ of torus $T^n$ acts on $\mathcal{H}$ as
\begin{equation}
    \mathcal{H}\rightarrow \mathcal{M}^{-T}\mathcal{H}\mathcal{M}^{-1}\,.
\end{equation}
For the representation \eqref{Htaurho} of moduli $\tau$ and $\rho$, the generic monodromy matrices can be parametrized as
\begin{equation}
\mathcal{M}_{\tau}=\left(\begin{array}{cc}
a&b\\
c&d
\end{array}\right),\quad
\mathcal{M}_{\rho}=\left(\begin{array}{cc}
a'&b'\\
c'&d'
\end{array}\right),
\end{equation}
with $ad-bc=1$ and $a'd'-b'c'=1$. Under the same representation as \eqref{H2}, the combined monodromy matrix is then constructed as
\begin{equation}\label{MSO22}
\mathcal{M}=\left(\begin{array}{cccc}
aa'&ba'&-bb'&ab'\\
ca'&da'&-db'&cb'\\
-cc'&-dc'&dd'&-cd'\\
ac'&bc'&-bd'&ad'
\end{array}\right)\in\mathrm{O}(2,2;\mathbb{Z})\,.
\end{equation}

The twist group $\mathrm{O}(n,n;\mathbb{Z})$ is generated by three fundamental types of global transformations \cite{Giveon:1988tt,Shapere:1988zv,Plauschinn:2018wbo}:
\begin{itemize}
    \item Diffeomorphism duality $\mathcal{M}_A$:
    \begin{equation}
 \mathcal{M}_A =  \left(\begin{array}{cc}
A^{-1} & 0 \\
0 & A^{T}
\end{array}\right),\quad A \in \mathrm{GL}(n,\mathbb{Z})\,,
\end{equation}
acting on the background geometry as a diffeomorphism $g+b\rightarrow A^T(g+b)A$;
\item $B$-field transformation $\mathcal{M}_B$:
\begin{equation}
\mathcal{M}_B =  \left(\begin{array}{cc}
\mathbbm{1} & 0 \\
B & \mathbbm{1}
\end{array}\right),\quad \text{with } B \text{ anti-symmetric},
\end{equation}
corresponding to a shift of the Kalb--Ramond field: $g\rightarrow g,\,b\rightarrow b+\alpha' B$;
\item Factorized dualities $\mathcal{M}_{\pm i}$:
\begin{equation}
 \mathcal{M}_{\pm i} =  \left(\begin{array}{cc}
\mathbbm{1}-E_i & \pm E_i \\
\pm E_i & \mathbbm{1}-E_i
\end{array}\right),
\end{equation}
where $E_i = \diag (0,\cdots,1,\cdots,0)$, with the 1 in the $i$-th position. This transformation corresponds to performing a duality along the $z^i$-circle.
\end{itemize}

Combining $\mathcal{M}_B$ with factorized dualities yields another useful transformation, which is called $\beta$-transformation:
\begin{equation}
 \mathcal{M}_\beta =  \left(\prod_{i=1}^n \mathcal{M}_{\pm i} \right)\mathcal{M}_{B(\beta)} \left(\prod_{i=1}^n \mathcal{M}_{\pm i}\right)=  \left(\begin{array}{cc}
\mathbbm{1} & \beta \\
0 & \mathbbm{1}
\end{array}\right),\quad \text{with } \beta \text{ anti-symmetric}.
\end{equation}

Back to our specific $T^2\times S^1$ model, we can select duality twists from the aforementioned transformations. These twists deform the geometry of the internal space and modify the periodicity conditions of the fields, thereby generating various generalized fluxes, including both geometric and non-geometric fluxes. Consider a twist $\mathcal{M}_{B(h)}$ defined as 
\begin{equation}
\mathcal{M}_{B(h)} =  \left(\begin{array}{cc}
\mathbbm{1} & 0 \\
B(h) & \mathbbm{1}
\end{array}\right),\quad \text{with }
B(h)=\left(\begin{array}{cc}
0&h\\
-h&0
\end{array}\right),\quad h\in\mathbb{Z}\,.
\end{equation}
This transformation is equivalent to introducing a constant $H$-flux, $H = \frac{\alpha'h}{2\pi} \ \mathrm{d} z_1\wedge\mathrm{d} z_2\wedge \mathrm{d} y$. The corresponding Kalb--Ramond field can be represented as
\begin{equation}
b=\left(\begin{array}{ccc}
0 & \frac{\alpha'hy}{2\pi} &0\\
-\frac{\alpha'hy}{2\pi} & 0 &0\\
0&0&0
\end{array}\right).
\end{equation}
Under this twist, the periodicity condition of the generalized metric becomes
\begin{equation}
\mathcal{H}(x+2\pi) = \mathcal{M}_{B(h)}^{-T}\mathcal{H}(x)\mathcal{M}_{B(h)}^{-1}\,.
\end{equation}
Performing a T-duality transformation $\mathcal{M}_{+1}$ along $z_1$-direction, the $B$-field gets trivial and the geometric $f$-flux is generated. Performing an additional T-duality transformation $\mathcal{M}_{+2}$ along $z_2$-direction, the $f$-flux is converted to $Q$-flux \cite{Kachru:2002sk}. The resulting internal space with non-trivial $Q$-flux forms a T-fold, which is locally geometric but globally non-geometric, because the transition function is the combination of diffeomorphism and T-duality \cite{ChristoferM.Hull_2005}. 

Now consider that over the internal space there is a duality twist
\begin{equation}\label{MBAbeta}
\mathcal{M} = \mathcal{M}_{B(h)} \mathcal{M}_{A(f)} \mathcal{M}_{\beta(q)}\,,
\end{equation}
with
\begin{equation}
A(f) =\left(\begin{array}{cc}
1 & -f \\
0 & 1
\end{array}\right),\quad \beta(q) =\left(\begin{array}{cc}
0 & q \\
-q & 0
\end{array}\right).
\end{equation}
Such a twist corresponds to a T-fold background with $H$-, $f$-, and $Q$-fluxes turned on, with flux numbers $h,\,f,\,q\in \mathbb{Z}$ respectively. Combining \eqref{MBAbeta} with the generic expression \eqref{MSO22}, we observe that the decomposed monodromy matrices for the two moduli are given by
\begin{equation}\label{Mtaurho}
\mathcal{M}_\tau =\left(\begin{array}{cc}
1 & f \\
0 & 1
\end{array}\right),\quad \mathcal{M}_\rho =\left(\begin{array}{cc}
1 & q \\
-h & 1-hq
\end{array}\right).
\end{equation}
The monodromy of the complex structure modulus $\tau$ naturally falls into the parabolic conjugacy class, whereas the $\rho$-monodromy can belong to either a parabolic, elliptic, or hyperbolic conjugacy class. Depending on the values of $h$ and $q$, $\operatorname{Tr}M_\rho^2$ takes different values and the $\rho$-monodromy can be classified as follows:
\begin{itemize}
    \item if one of $h$ and $q=0$, the monodromy is a shift on $\rho$ and is parabolic, similar as $\mathcal{M}_\tau$;
    \item if $hq=1$, it is a $\mathbb{Z}_6$ elliptic monodromy. For $h=q=1$ it is indeed $\mathcal{M}_6$ in \eqref{M2346};
    \item if $hq=2$, it is a $\mathbb{Z}_4$ elliptic monodromy, which is conjugate to $\mathcal{M}_4$ in \eqref{M2346} by $\SLtwoZ$-conjugation, but the Minkowski minimum situation is different. For example, for $h=2$, $q=1$,
    \begin{equation}
        \mathcal{M}_\rho = \left(\begin{array}{cc}
1 & 1 \\
-2 & -1
\end{array}\right) = \left(\begin{array}{cc}
1 & 0 \\
-1 & 1
\end{array}\right) \mathcal{M}_4 \left(\begin{array}{cc}
1 & 0 \\
-1 & 1
\end{array}\right)^{-1},
    \end{equation}
    and the minimum is at $\rho = \frac{-1+i}{2}$;
    \item if $hq=3$, it is a $\mathbb{Z}_3$ elliptic monodromy conjugate to $\mathcal{M}_3$;
    \item if $hq<0$, the monodromy is hyperbolic and conjugate to $\mathcal{M}_h$;
    \item if $hq\geq 4$, the monodromy $M_\rho$ becomes imaginary, corresponding to complexified generators of the duality group. For $hq>4$, the monodromy is hyperbolic and the mass matrix $M_\rho$ is
    \begin{equation}
       \left(\begin{array}{cc}
            \sqrt{\frac{hq}{hq-4}}\arctan\left[-\sqrt{1-\frac{4}{(hq-2)^2}}\right] & 2\sqrt{\frac{q}{h(hq-4)}}\arctan\left[-\sqrt{1-\frac{4}{(hq-2)^2}}\right] \\
            2\sqrt{\frac{h}{q(hq-4)}}\arctan\sqrt{1-\frac{4}{(hq-2)^2}} & \sqrt{\frac{hq}{hq-4}}\arctan\sqrt{1-\frac{4}{(hq-2)^2}}
        \end{array}\right)+i\pi\left(\begin{array}{cc}
            1 &  \\
             & 1
        \end{array}\right).
    \end{equation}
    This gives a minimal orbit at $h|\rho|^2+hq\rho_1+q=0$, with $\rho_{2,\,\max}=\sqrt{\frac{q}{h}}\frac{\sqrt{hq-4}}{2}$. At the orbit the potential picks the minimal value, with
    \begin{equation}
    v^\rho_{\min}\equiv4\left(\log\frac{hq-2-\sqrt{hq(hq-4)}}{2}\right)^2-4\pi^2\,,
    \end{equation}
    contributing to the potential 
    \begin{equation}
        V\equiv 2e^{2(\alpha-\beta)\phi}\left(v^\tau+v^\rho\right).
    \end{equation}
    Such a minimum is negative for $hq\leq 25$ ($\operatorname{Tr}M_\rho^2<0$) and positive for $hq > 25$ ($\operatorname{Tr}M_\rho^2>0$). 
    
    For $hq=4$, it is parabolic. As an example, for $h=q=2$,
    \begin{equation}
        M_\rho = \left(\begin{array}{cc}
            -2+i\pi & -2 \\
            2 & 2+i\pi
        \end{array}\right).
    \end{equation}
    There is a minimum at $\rho_1=-1,\,\rho_2\to 0$, with value 
    \begin{equation}
    v^\rho_{\min}\equiv\operatorname{Tr}\left(M_\rho^2+M^T_\rho \mathcal{H}^{-1}_\rho M_\rho \mathcal{H}_\rho\right)_{\min} \to -4\pi^2\,.
    \end{equation}
    Conclusively, stabilized $\rho$ with $hq\geq 4$ gives a non-zero constant contribution to the potential and dominates over the runaway $\tau$ contribution. For $4\leq hq\leq 25$, the potential $V$ negatively diverges when it goes towards $\phi\to -\infty$, where the Scherk--Schwarz radius shrinks to a point and the effective theory breaks down. For $hq>25$, the modulus $\rho$ is stabilized at the valley of the potential, while the radion remains a runaway direction along which the potential tends to zero. Additional fluxes and non-perturbative effects \cite{Quiros:2004fn,Parameswaran:2024mrc} may stabilize the Scherk--Schwarz radion to create a potential minimum.
\end{itemize}
In addition, the $\mathbb{Z}_2$ monodromy can be achieved via a further T-duality such that $\mathcal{M}_\rho\to-\mathcal{M}_\rho$.

\subsection{Dark Dimension realization in the toy model}
We now endeavor to realize the Dark Dimension Scenario within the framework of our seven-dimensional toy model. Referring to \eqref{Mtaurho}, we assign a parabolic monodromy for $\tau$ and an elliptic monodromy for $\rho$. For instance, by activating fluxes with $h=2,\,q=1$ and $f$ arbitrary, the monodromy generator $M$ is given by
\begin{equation}\label{Mp4}
M_{p,4}=\left(\begin{array}{cccc}
\frac{\pi}{2}&f&0&\frac{\pi}{2}\\
0&\frac{\pi}{2}&-\frac{\pi}{2}&0\\
0&\pi&-\frac{\pi}{2}&0\\
-\pi&0&-f&-\frac{\pi}{2}
\end{array}\right).
\end{equation}
Incorporating this monodromy generator with the moduli matrix defined in equation \eqref{H2}, the scalar potential \eqref{V} becomes 
\begin{equation}\label{Vp4}
V= 2e^{2(\alpha-\beta)\phi} \left(\frac{f^2}{\tau_2^2} + \frac{\pi^2}{4}\frac{4|\rho|^4+4\rho_1\left(2|\rho|^2+2\rho_1+1\right)+1}{\rho_2^2}\right) \overset{\rho \text{ stab.}}{=\joinrel=} 2e^{2(\alpha-\beta)\phi}\frac{f^2}{\tau_2^2}\,.
\end{equation}
In this expression, the K\"ahler scalar $\rho$ of this potential is stabilized at $\rho=\frac{-1+i}{2}$. The real part $\tau_1$ is a flat direction, while the imaginary part $\tau_2$ exhibits runaway behavior, driving the system towards a large complex structure limit. To further analyze the potential, we perform a reparameterization of the scalar fields as follows:
\begin{equation}
e^{a\phi} =x = r\cos\vartheta,\quad \tau_2^2 = y = r\sin\vartheta ,\,\quad a = 2(\beta-\alpha),
\end{equation}
with $0<\vartheta<\frac{\pi}{2}$. Hence, we can express the potential in terms of a trajectory field $r$ and a transverse field $\vartheta$. Substituting these variables into the potential, we obtain:
\begin{equation}\label{Vp2}
V = \frac{2f^2}{r^2}\frac{1}{\cos\vartheta\sin\vartheta}\,,
\end{equation}
which stabilizes the modulus $\vartheta$ at $\vartheta = \pi/4$, corresponding to the opposite direction of the gradient of the potential. Consequently, as the system evolves over time, the potential approaches:
\begin{equation}\label{Vpstab}
V_{\text{stab.}} = \frac{4f^2}{r^2} = 4f^2 e^{-2a\phi} \propto m_{\text{KK}}^4\,.
\end{equation}
This result indeed satisfies the requirements of the Dark Dimension Scenario, where the dark dimension is naturally identified with the Scherk--Schwarz $S^1$-radius. We will comment on this potential in the next section, where the 4-dimensional result can be similar to \eqref{Vp4} by stabilizing more moduli.

\section{Dark Dimension realization in 4-dimension}\label{sec5}
The method used in the toy model can be naturally extended to 4-dimension. Consider a type II superstring theory compactified on $T^5\times S^1$. We perform a Scherk--Schwarz reduction over $S^1$, with twists taken from the T-duality group $\mathrm{O}(5,5;\mathbb{Z})$ of $T^5$. We truncate the Ramond-Ramond fields and focus on the NS-NS background moduli $\mathcal{H}$, as given by expression \eqref{H}, to fit into the representation of the T-duality group. As in the toy model, we require at least one runaway scalar from $\mathcal{H}$, while the other scalars are either flat or stabilized at Minkowski minima. In the last section, we utilize the parabolic conjugacy class of $\SLtwoZ$ to generate the required runaway potential, and the elliptic conjugacy class to stabilize the $T^2$-volume at $V^\rho=0$. However, the conjugacy structures of larger groups are highly complex; for example, there is still no explicit characterization of the conjugacy classes of $\mathrm{SL}(3,\mathbb{Z})$ \cite{Achmed_Zade_2018}.

Nevertheless, if we set aside the moduli stabilization problem, the toy model provides a straightforward example of duality twists. The torus $T^5$ can be decomposed as $T^2_{A}\times T^2_{B}\times S^1$, with coordinates $\{z^1,z^2;z^3,z^4;z^5\}$. We can turn on the T-fold fluxes with corresponding monodromy given by \eqref{Mtaurho} for each $T^2$ over the base $S^1$. As an example, for both $T^2$'s we take the monodromy as in \eqref{Mp4}. Then the potential \eqref{V} is
\begin{equation}
    \begin{aligned}
        V = 2e^{2(\alpha-\beta)\phi} &\left(\frac{f_{A}^2}{\left(\tau_2^{A}\right)^2} + \frac{\pi^2}{4}\frac{4|\rho^{A}|^4+4\rho_1\left(2|\rho^{A}|^2+2\rho^{A}_1+1\right)+1}{\left(\rho^{A}_2\right)^2}\right.\\
        &\left.  +\frac{f_{B}^2}{\left(\tau_2^{B}\right)^2} + \frac{\pi^2}{4}\frac{4|\rho^{B}|^4+4\rho_1\left(2|\rho^{B}|^2+2\rho^{B}_1+1\right)+1}{\left(\rho^{B}_2\right)^2}\right),
    \end{aligned}
\end{equation}
where $2(\alpha-\beta)=-\sqrt{3}$. Upon stabilizing $\rho^{A}$ and $\rho^{B}$ as in \eqref{Mp4}, the potential simplifies to
\begin{equation}
V = 2e^{-\sqrt{3}\phi}\left(\frac{f_{A}^2}{\left(\tau_2^{A}\right)^2}+\frac{f_{B}^2}{\left(\tau_2^{B}\right)^2}\right).
\end{equation}
Similar to \eqref{Vp2}, the potential can be further stabilized by reparametrizing the remaining moduli into spherical coordinates
\begin{equation}
e^{\sqrt{3}\phi} = r\cos\vartheta,\quad \left(\tau_2^{A}\right)^2 = r\sin\vartheta\cos\varphi,\quad\left(\tau_2^{B}\right)^2 = r\sin\vartheta\sin\varphi\,,
\end{equation}
with $0<\vartheta,\varphi<\frac{\pi}{2}$. The potential now becomes
\begin{equation}
V = \frac{2}{r^2\cos\vartheta\sin\vartheta}\left(\frac{f_{A}^2}{\cos\varphi}+\frac{f_{B}^2}{\sin\varphi}\right),
\end{equation}
which attains a minimum at $\vartheta=\pi/4$, $\tan\varphi = \left(f_{B}/f_{A}\right)^{\frac{2}{3}}$, where
\begin{equation}\label{Vf1234}
V = \frac{4}{r^2}\left(f_{A}^{\frac{4}{3}}+f_{B}^{\frac{4}{3}}\right)^{\frac{3}{2}} = 2e^{-2\sqrt{3}\phi}\left(f_{A}^{\frac{4}{3}}+f_{B}^{\frac{4}{3}}\right)^{\frac{3}{2}} \propto m_{\text{KK}}^4\,.
\end{equation}

In this model, we indeed obtain the Dark Dimension power relation between the potential and the $S^1_y$-Kaluza--Klein scale. The complexified volumes of the two subtori are fixed, and two shape scalars run away. However, there remain many flat directions in the field space, especially since the $z^5$-direction is not stabilized. In the Dark Dimension scenario, the number of large internal dimensions should be one, and an additional uncontrolled radius could be problematic.

On the other hand, stabilizing all relevant moduli in $\mathcal{H}$ via nontrivial monodromies in $\mathrm{O}(5,5;\mathbb{Z})$ is almost impossible. Here we choose to keep some off-diagonal metric fields and corresponding $B$-field components flat, and decompose the twist group to $\mathrm{O}(3,3;\mathbb{Z})\times \mathrm{O}(2,2;\mathbb{Z})$ of $T^3\times T^2$. For the twist over $T^3$, we only turn on $H$- and $Q$-fluxes. 

Denote the coordinates of $T^3$ to be $\left\{w^1,w^2,w^3\right\}$. Note that we can take a monodromy $\mathcal{M}_B$ such that the Kalb--Ramond field over $T^3$ is
\begin{equation}
b_{ij}= -b_{ji} = \frac{\alpha'h_{ij}y}{2\pi} ,\quad i>j,\,\,i,j=1,2,3,
\end{equation}
where $h_{ij}$ are the $H$-flux numbers. This configuration yields a constant $H$-flux over $T^3\times S^1$. Applying factorized T-duality transformations $\mathcal{M}_{+j}\mathcal{M}_{+i}$ on this background, the twist becomes a $\beta$-transformation $\mathcal{M}_\beta$, converting the $H$-flux components into corresponding $Q$-fluxes. Now consider a monodromy $\mathcal{M}=\mathcal{M}_{B(h)}\mathcal{M}_{\beta(q)}\in \mathrm{O}(3,3;\mathbb{Z})$, such that
\begin{equation}\label{M33}
\mathcal{M}=\begin{pmatrix}
1&0&0&0&q_{12}&q_{13}\\
0&1&0&-q_{12}&0&q_{23}\\
0&0&1&-q_{13}&-q_{23}&0\\
0&h_{12}&h_{13}&1-h_{12}q_{12}-h_{13}q_{13}&-h_{13}q_{23}&h_{12}q_{23}\\
-h_{12}&0&h_{23}&-h_{23}q_{13}&1-h_{12}q_{12}-h_{23}q_{23}&-h_{12}q_{13}\\
-h_{13}&-h_{23}&0&h_{23}q_{12}&-h_{13}q_{12}&1-h_{13}q_{13}-h_{23}q_{23}
\end{pmatrix},
\end{equation}
with flux numbers $h_{ij},q_{ij}\in\mathbb{Z}$. The indices of flux numbers are anti-symmetric: $h_{ji}=-h_{ij}$, $q_{ji}=-q_{ij}$.

The conjugacy classes of $\mathcal{M}$ remain complicated. We require that the volume of $T^3$ is stabilized at a Minkowski minimum. To achieve that, $\mathcal{M}$ should be conjugated to a rotation. Suppose all flux numbers are nontrivial, then numerical cases indicate that the only possible values might be $h_{12}q_{12}=h_{13}q_{13}=h_{23}q_{23}=1$. As an example, we choose $h_{12}=q_{12}=h_{13}=q_{13}=1$, $h_{23}=q_{23}=-1$. This configuration has a minimum $V_{T^3}=0$ at
\begin{equation}
b_{12}=-b_{13}=b_{23}=-\frac{1}{2},\quad g_{ij} = \frac{1}{2}\delta_{ij}\,.
\end{equation}
For the twist over $T^2$, we take the monodromy \eqref{MBAbeta} with generator \eqref{Mp4} as an example, such that its K\"ahler modulus is stabilized by $H$- and $Q$-fluxes, and the complex structure modulus runs away. Then the potential \eqref{V} becomes
\begin{equation}\label{VT5}
V = V_{T^3}+V_{T^2} = V_{T^2} = 2e^{-\sqrt{3}\phi} \left(\frac{f^2}{\tau_2^2} + \frac{\pi^2}{4}\frac{4|\rho|^4+4\rho_1\left(2|\rho|^2+2\rho_1+1\right)+1}{\rho_2^2}\right),
\end{equation}
which is the same as \eqref{Vp4}. After evolving for enough time, the potential becomes proportional to $m_{\text{KK}}^4$, aligning with the requirements of the Dark Dimension Scenario. 

The flux construction presented here, featuring a duality-twisted $T^5$ over a Scherk--Schwarz $S^1$ base, inherently satisfies the Bianchi identities \cite{Shelton:2005cf}. The NS-NS Bianchi identities, which arise from applying T-duality to the condition $\mathrm{d}H=0$, impose constraints on the allowed values of the geometric and non-geometric fluxes. In our construction, which is free of $R$-flux and R-R fluxes, the relevant Bianchi identities for the non-vanishing $H$-, $f$-, and $Q$-fluxes are expressed as:
\begin{align}
    H_{e[ab}{F^e}_{cd]}&=0\,,\label{BI1}\\
    {F^a}_{e[b}{F^e}_{cd]}+H_{e[bc}{Q^{ae}}_{d]}&=0\,,\\
    {Q^{ab}}_e {F^e}_{cd}-4{F^{[a}}_{e[c}{Q^{c]e}}_{d]}&=0\,,\\
    {Q^{[ab}}_e {Q^{c]e}}_d&=0\,.
\end{align}
To map these general expressions to our specific flux components, we first denote the $T^2$ coordinates as $\left\{w^4,w^5\right\}$, such that $h\equiv h_{45}$, $f\equiv f_{45}$, $q\equiv q_{45}$. The $H$-flux components are then identified as $H_{aby}\equiv h_{ab}$, where $a,b=1,2,3,4,5$ and $y$ denotes the Scherk--Schwarz $S^1$ direction. Similarly, the $Q$-flux components are identified as ${Q^{ab}}_{y}\equiv q_{ab}$, satisfying upper-indices anti-symmetry ${Q^{ab}}_{c}=-{Q^{ba}}_{c}$. Given that the $f$-flux vanishes on the $T^3$ subspace, its only non-trivial components are ${F^1}_{2y}=-{F^1}_{y2}\equiv f$. A direct computation confirms that all Bianchi identities are satisfied, both for components purely on the $T^2$ and $T^3$ subspaces and for components involving their intersection. This result holds primarily because the T-duality transformations defining our twist in \eqref{Mtaurho} and \eqref{M33} do not act on the Scherk--Schwarz circle, which ensures the absence of any ${F^y}_{ab}$ or ${Q^{ya}}_b$ flux components. Furthermore, the first identity, \eqref{BI1}, provides a consistency check that justifies our decomposition $T^5 = T^3\times T^2$.

Our model yields an effective theory characterized by a constant $T^5$-volume and an expanding Scherk--Schwarz $S^1$-radius. The additional runaway scalar is the complex structure of some $T^2$ subtorus within $T^5$, and the volume of this $T^2$ is fixed. Consequently, for a square torus, this dynamic means one direction of the $T^2$ shrinks while the other expands. Denote the torus radii as $R_1$ and $R_2$. The imaginary part of the complex structure modulus is then given by $\tau_2= R_1/R_2$. The potential \eqref{Vp2} is stabilized at $\vartheta=\pi/4$, while along this runaway direction, the Scherk--Schwarz radius satisifies $\varrho = e^{\sqrt{3}\phi}=\tau_2$. Since the growth of $\tau_2$ is sourced by both an increasing $R_1$ and a decreasing $R_2$, it follows that the expansion rate of $R_1$ is necessarily slower than the growth rate of $\varrho$. After long-time evolution, this results in a significantly larger Scherk--Schwarz radius, identifying it as the dark dimension.

In our model, supersymmetry is completely broken due to the runaway potential, with the supersymmetry-breaking scale at least the scale of the gravitino mass. The gravitini belong to the representation of the $R$-symmetry group, which is the compact subgroup of the T-duality associated with rotational (elliptic) conjugacy classes. By applying a rotational twist, we simultaneously stabilize the volume and the $B$-field and impart mass to the gravitini, as elaborated in the seminal work by Scherk and Schwarz \cite{Scherk:1979zr}. In the string frame, the gravitino mass is inversely proportional to $\varrho$ \cite{Kiritsis:1996xd}, such that in the Einstein frame, 
\begin{equation}
M_{3/2} \sim \varrho^{-\frac{3}{2}} = e^{-\frac{3}{2}\beta\phi} = m_{\text{KK}}\,,
\end{equation}
which is much higher than the scale of our potential.

Furthermore, if $f=0$ such that the monodromy is purely elliptic ($V_{\text{classical}}=0$), the theory allows for an exact worldsheet description and features an asymmetric orbifold internal space \cite{Condeescu:2012sp,Gkountoumis:2023fym}, and the 1-loop Casimir potential of 4D Scherk--Schwarz compactifications is known as \cite{Abel:2016hgy,Parameswaran:2024mrc}
\begin{equation}\label{V1loop}
V_{\text{Casimir}}=\xi \left(N^0_\text{f}-N^0_\text{b}\right)m_{\text{KK}}^4\,,
\end{equation}
where $\xi$ is an $\mathcal{O}(10^{-3})$ number \cite{Abel:2016hgy} and $N^0_\text{f}-N^0_\text{b}\sim\mathcal{O}(10)$ \cite{Parameswaran:2024mrc}. In this case, more massless bosons than massless fermions exist around the Minkowski vacuum from the elliptic monodromy, resulting in a negative one-loop Casimir energy. This energy becomes unstable as $\phi\to-\infty$, corresponding to the zero-volume and strong coupling limit. Stabilization of this potential might be achievable by introducing proper fluxes \cite{Parameswaran:2024mrc}. However, in our models with non-trivial $f$-fluxes and lacking a worldsheet description, the 1-loop Casimir correction for a twisted background with a runaway potential remains undetermined, and the mass spectrum is time-dependent. Suppose the quantum corrections in our model mimic the expression in \eqref{V1loop}, it follows that the Casimir corrections are strongly suppressed relative to the classical Scherk--Schwarz potential, $V_{\text{Casimir}} \ll V_{\text{classical}}$. This suppression occurs because the coefficient of the classical potential $4f^2\gtrsim\mathcal{O}(1-10)$ for proper flux numbers values, making it substantially larger than the factor $\xi\left(N^0_\text{f}-N^0_\text{b}\right)\sim-\mathcal{O}(10^{-2})$ that governs the magnitude of the Casimir energy.

\section{Conclusion and Outlook}
In this work, we presented a novel approach to integrating the concept of the Dark Dimension with non-geometric compactifications. Our investigation demonstrates that the Dark Dimension Scenario can be realized within the framework of type II superstring theory by employing a Scherk--Schwarz reduction mechanism combined with T-folds. This T-fold compactification on $T^5\times S^1$ utilizes $T^5$ T-duality $\mathrm{O}(5,5;\mathbb{Z})$ as the twist group, and $S^1$ serves as the Scherk--Schwarz radius. 

To construct a Scherk--Schwarz potential with the required properties, the duality twist over $T^5$ was determined. We decomposed the T-duality to be the subgroup $\mathrm{O}(2,2;\mathbb{Z})\times \mathrm{O}(3,3;\mathbb{Z})$. In the $T^2$ segment, characterized by T-duality $\mathrm{O}(2,2;\mathbb{Z}) \sim \SLtwoZ_\tau \times \SLtwoZ_\rho$, we engineered twists to ensure the complex structure modulus $\tau$ experiences a parabolic monodromy, while the Kähler modulus $\rho$ undergoes an elliptic monodromy. This parabolic monodromy of $\tau$ is characterized by the $f$-flux, and the monodromy of $\tau$ is characterized by $H$- and $Q$-flux. We also provided a comprehensive classification of $\SLtwoZ$ conjugacy classes in terms of $H$-, $Q$-flux numbers in Subsec.~\ref{hq}, especially noting the imaginary monodromy with $hq\geq 4$. For the remaining $T^3$ component, we identified elliptic conjugacy classes of $\mathrm{O}(3,3;\mathbb{Z})$ like \eqref{M33}. The elliptic conjugacy classes stabilize corresponding moduli and contribute $V_e=0$ to the potential, whereas the parabolic monodromy in our setup results in a runaway potential for $\tau_2$. By stabilizing the turning to another runaway direction $\phi$, we aligned these two scalars to produce a potential that towards a desirable direction with $V\propto e^{-2\sqrt{3}\phi}\propto m_{\text{KK}}^4$, as required by the Dark Dimension.

Note that the moduli stabilization remains challenging. With our decomposition of T-duality into $\mathrm{O}(3,3;\mathbb{Z})\times \mathrm{O}(2,2;\mathbb{Z})$, some off-diagonal scalar fields present flat directions. Further exploration into conjugacy classification could potentially stabilize more moduli. Investigating the tadpole contributions and Bianchi identities may allow for the stabilization of all NS-NS moduli using non-geometric fluxes, as discussed in \cite{Plauschinn:2020ram}. Additional moduli might be stabilized by RR fluxes. However, our model necessitates that the Scherk--Schwarz radion is not stabilized and at least one more modulus contributes to a runaway potential. Fully understanding the behaviors of moduli under these conditions requires quantum corrections at various levels, which are complex to compute in a dynamic background. Future research may address these challenges.

We achieved the required exponent by aligning other scalar fields with the radion. Dynamically, the scalars can move along a non-geodesic trajectory before approaching the steepest direction. Suppose that the current universe has the potential proportional behavior in the Dark Dimension Scenario, but is not near the boundary of the field space, then the potential after angular stabilization could run faster than $m^4_{\text{KK}}$. The multi-field quintessence models follow the same idea to acquire the transient cosmic acceleration. However, within our framework, the exponent in the expression $V \propto e^{-2\sqrt{3}\phi}$, indicative of an asymptotic exponential quintessence model, is too large to feasibly achieve the cosmic acceleration. For additional discussion, see Appendix~\ref{AppB}.

\section*{Acknowledgements}
We thank George Gkountoumis, Thomas Grimm, Miguel Montero, and Irene Valenzuela for the useful discussions, and thank Ivano Basile and Dieter L\"ust for the discussions after the first version. The work of G.N. is supported by the China Scholarship Council.

\appendix
\section{$T^2\times S^1$ U-fold compactification}\label{AppA}
Consider a compactification of type IIB superstring theory on $T^2\times S^1$ or an 11D supergravity on $T^3\times S^1$. We implement a Scherk--Schwarz reduction over the $S^1$, with the monodromy matrix $\mathcal{M}=e^M$ a component of the U-duality group of $T^2$. The U-duality group is isomorphic to $\SLtwoZ\times\mathrm{SL}(3,\mathbb{Z})$ \cite{Hull:1994ys}. Within this setup, the theory comprises a gravity multiplet that includes seven scalars: the K\"ahler modulus $\rho$, the complex structure modulus $\tau$, the axio-dilaton $\sigma=c_0+ie^{-\phi_{10}}$, and the reduced RR 2-form $c_2$. The corresponding moduli space is described by
\begin{equation}
\frac{\mathrm{SL}(2,\mathbb{R})}{\SLtwoZ\times \mathrm{U}(1)}\times\frac{\mathrm{SL}(3,\mathbb{R})}{\mathrm{SL}(3,\mathbb{Z})\times \mathrm{SO}(3)}\,.
\end{equation}
We organize the scalar fields into the adjoint representation $\mathcal{H}\in \mathrm{SL}(2,\mathbb{R})_\tau\times \mathrm{SL}(3,\mathbb{R})$. In this structure, the complex structure modulus $\tau$ transforms under $\mathrm{SL}(2,\mathbb{R})\tau$, while other fields fall under $\mathrm{SL}(3,\mathbb{R})$. The composition of the fields in $\mathrm{SL}(3,\mathbb{R})$ is given by \cite{Liu_1998,Castellano:2023aum}
\begin{equation}
    \mathcal{H}_{\mathrm{SL}(3)} = \frac{e^{\phi_8/3}}{\rho_2}\left(\begin{array}{ccc}
     \rho_2 e^{-\phi_8}+ |c_2+  \sigma_1 \rho|^2  & c_2\rho_1+\sigma_1 |\rho|^2 & c_2+\sigma_1 \rho_1 \\
     c_2\rho_1+\sigma_1 |\rho|^2  & |\rho|^2 &  \rho_1 \\
     c_2+\sigma_1 \rho_1  & \rho_1 & 1
    \end{array}\right).
\end{equation}
Despite the complexity of the complete conjugacy classification of $\mathrm{SL}(3,\mathbb{Z})$, stabilizing the scalars requires only a suitable elliptic conjugacy class. Here, we select a $\mathbb{Z}_2$-monodromy
\begin{equation}
    \mathcal{M}_{\mathrm{SL}(3)} = e^M = \left(\begin{array}{ccc}
     0  & 0 & 1 \\
     0  & -1 &  0 \\
     1  & 0 & 0
    \end{array}\right), \quad \text{with } M= \frac{\pi}{\sqrt{2}}\left(\begin{array}{ccc}
     0  & 1 & 0 \\
     -1  & 0 &  1 \\
     0  & -1 & 0
    \end{array}\right).
\end{equation}
This configuration allows the $\mathrm{SL}(3)$ fields to contribute to a Minkowski minimum, effectively stabilizing the moduli at $\rho=i,\,\sigma=i,\,c_2=0$. For the complex structure modulus $\tau$, we apply the parabolic conjugacy class of $\mathrm{SL}(2,\mathbb{R})_\tau$ as the monodromy. Then we get the same potential as \eqref{Vp4} and so on in our T-fold constructions.

\section{Multi-field quintessence}\label{AppB}
The runaway potentials discussed in this paper are characteristic examples of multi-field quintessence models. Consider a single-field quintessence model in a flat space with zero spatial curvature. The potential is given by:
\begin{equation}
V = V_0\ e^{-\lambda\phi}\,,
\end{equation}
where to sustain an eternally accelerating universe, the exponent $\lambda$ must satisfy $\lambda \leq \sqrt{2}$. Asymptotically, quintessence models incorporating multiple fields typically exhibit trajectories with steeper declines. For example, a potential involving two fields might be expressed as
\begin{equation}
V = V_0\ e^{-\lambda\left(\phi_1+\phi_2\right)}\,,
\end{equation}
with $\phi_1,\phi_2$ canonically normalized. By naively defining a new normalized field $\Phi = \frac{1}{\sqrt{2}}(\phi_1 + \phi_2)$, we observe that the stabilized trajectory aligns along $\Phi$, resulting in a potential of the form $V=V_0 e^{-\lambda_{\text{eff}}\Phi}$, with $\lambda_{\text{eff}}=\sqrt{2}\lambda$. 

However, depending on initial conditions, $\lambda_{\text{eff}}$ may vary over time. For example, if the initial universe was in a kination epoch before the 5D decompactification as described by \eqref{VT5} or \eqref{Vp4}, it would initially evolve along the $\vartheta$-direction, effectively setting $\lambda_{\text{eff}} = 0$. As the evolution progresses, $\lambda_{\text{eff}}$ increases until it aligns with the steepest descent. Note that in our models, the field space metric for $\tau$ is
\begin{equation}
\mathrm{d} s_\tau^2= \frac{4}{\tau_2^2}\left(\mathrm{d}\tau_1^2+\mathrm{d}\tau_2^2\right).
\end{equation}
This yields a maximal $\lambda_{\text{eff}}>2\sqrt{3}$, which is unsuitable for supporting the required accelerated expansion phase in our universe. However, with more runaway directions and slowly descending scalar fields we may realize the cosmic accelerated Dark Dimension. 

Furthermore, general theories might include couplings between the radion or dilaton $\phi$ and the kinetic terms of other fields. Consider the theory
\begin{equation}
\mathcal{L}_{\text{scalar}} = -\frac{1}{2} \partial^\mu \phi\partial_\mu \phi - \frac{1}{2}e^{-2k\phi} \partial^\mu \chi\partial_\mu \chi - e^{-\lambda_1\phi-\lambda_2\chi}\,.
\end{equation}
The kinetic term corresponds to an axion-saxion pair. The dynamics of such a theory is also discussed in for example \cite{Cicoli:2023opf}. Notably, for large values of $\phi$, the kinetic term of $\chi$ becomes exponentially suppressed. The normalization of the effective field $\Phi$ in $e^{-\lambda_{\text{eff}}\Phi} = e^{-\lambda_1\phi - \lambda_2\chi}$ is predominantly influenced by $\phi$, such that the value of $\lambda_{\text{eff}}$ is slightly elevated above $\lambda_1$. As detailed in \cite{Andriot:2024jsh}, the universe may experience a transient accelerating expansion if $\lambda < \sqrt{3} + \mathcal{O}(0.01)$ for a flat universe or $\lambda < \sqrt{3} + \mathcal{O}(0.1)$ for an open universe. By setting $\lambda_1 = \sqrt{3}$, as in our Scherk--Schwarz models, $\lambda_{\text{eff}}$ could evolve from zero to a value slightly exceeding $\lambda = \sqrt{3}$, thereby enabling an accelerated expansion history.

\newpage
\bibliographystyle{JHEP}
\bibliography{bib}

\end{document}